\documentclass[fleqn,usenatbib,letters]{mnras}

\usepackage{newtxtext,newtxmath}

\usepackage[T1]{fontenc}

\DeclareRobustCommand{\VAN}[3]{#2}
\let\VANthebibliography\thebibliography
\def\thebibliography{\DeclareRobustCommand{\VAN}[3]{##3}\VANthebibliography}

\usepackage{graphicx}	
\usepackage{amsmath}	
\usepackage{placeins}
\usepackage{xcolor}
\usepackage{multirow}
\usepackage{physics}
\usepackage{orcidlink}

\newcommand{\st}[1]{_\text{#1}}

\newcommand{\emax}{E_e^{\mathrm{max}}}
\newcommand{\emin}{E_e^{\mathrm{min}}}
\newcommand{\ecb}{E_e^{\mathrm{cb}}}

\newcommand{\Egcbobs}{E_{\gamma}^{\mathrm{b,obs}}}

\newcommand{\uvot}{\textit{Swift}-UVOT}
\newcommand{\xrt}{\textit{Swift}-XRT}
\newcommand{\bat}{\textit{Swift}-BAT}
\newcommand{\gbm}{\textit{Fermi}-GBM}
\newcommand{\lat}{\textit{Fermi}-LAT}

\newcommand{\lhaaso}{LHAASO}

\title[GRB~221009A - MWL Afterglow Spectrum]{The Multiwavelength Picture of GRB~221009A's Afterglow}

\author[M. Klinger et al.]{
Marc Klinger$^{1}$\thanks{E-mail: marc.klinger@desy.de}\orcidlink{0000-0002-4697-1465},
Andrew M. Taylor$^{1}$\orcidlink{0000-0001-9473-4758},
Tyler Parsotan$^{2}$\orcidlink{0000-0002-4299-2517},
Andrew Beardmore$^{3}$, 
Sebastian Heinz$^{4}$\orcidlink{0000-0002-8433-8652}
\newauthor
and Sylvia J. Zhu$^{1}$\orcidlink{0000-0002-6468-8292}\\
$^{1}$Deutsches Elektronen-Synchrotron DESY, Platanenallee 6, 15738 Zeuthen, Germany \\
$^{2}$Astrophysics Science Division, NASA Goddard Space Flight Center,Greenbelt, MD 20771, USA. \\
$^{3}$School of Physics and Astronomy, University of Leicester, Leicester LE1 7RH, UK\\
$^{4}$Department of Astronomy, University of Wisconsin-Madison, 475 N. Charter Street, Madison, WI 53706, USA.
}

\date{
Accepted 2023 November 30. Received 2023 November 30; in original form 2023 August 17}

\pubyear{2023}

\begin{document}
\label{firstpage}
\pagerange{\pageref{firstpage}--\pageref{lastpage}}
\maketitle

\begin{abstract}
We present counts-level fits to the multi-instrument (keV--GeV) data of the early afterglow (4~ks, 22~ks) of the brightest gamma-ray burst detected to date, GRB~221009A. 
The complexity of the data reduction, due to the unprecedented brightness and the location in the Galactic plane, is critically addressed. 
The energy spectrum is found to be well described by a smoothly broken power law with a break energy at a few keV. 
Three interpretations (slow/fast cooling or the transition between these) within the framework of forward shock synchrotron emission, from accelerated and subsequently cooled electrons, are found. The physical implications for each of these scenarios are discussed.
\end{abstract}




\section{Introduction} \label{sec:intro}
The non-thermal afterglow emission of gamma-ray bursts (GRBs) constitutes a two-dimensional flux surface as a function of energy and time.
In the energy dimension the cleanest observational window ranges from hard X-ray (keV) up to very-high-energy (VHE; >100 GeV) gamma-ray energies. In the VHE band and above, the absorption due to the extragalactic background light (EBL) limits spectral inference. From the soft X-ray down to infra-red bands, both photoelectric absorption and dust scattering also prohibit decisive conclusions, such that optical and UV data serve only as lower limits to the intrinsic source flux. We emphasise that the systematic uncertainty introduced by correcting for these exponential absorption effects is difficult to constrain. At even lower energies, i.e. from infra-red down to the radio band, the propagation becomes unhindered once again; however, an additional emission component, usually attributed to the reverse shock, appears to begin to dominate within this range, limiting the insight on the forward shock component. 
In the dimension of time, the afterglow's most informative data window lies at early times, starting with the clear onset of the afterglow. Towards later times the significance of the GRB signal over the background quickly drops due to the decaying light curves in combination with the limited sensitivity of GRB detection instruments. 

GRB~221009A is the brightest gamma-ray burst detected to date. Its extensive multi-wavelength coverage promises a correspondingly large amount of new insights. Contrary to these expectations, the extraordinarily high flux and its location in the Galactic plane (latitude b = 4.3 degrees) led to additional systematic errors in several instruments, hindering the measurement of the GRB's flux. This is not only limited to the prompt emission phase ($\lesssim T_{0\mathrm{,GBM}} + 700$~s, \citealt{FermiGBM_grb221009a}); for example its complex dust echos significantly challenge the determination of the afterglow X-ray spectrum up to more than 100~ks \citep[e.g.][]{swift_grb221009a}.

We focus thus on the early afterglow with contemporaneous data from keV up to GeV energies, in particular two time intervals around 4~ks and 22~ks after the trigger time of the \textit{Fermi} Gamma-Ray Burst Monitor ({\gbm}). In both intervals the \textit{Swift} X-ray Telescope ({\xrt}; \citealt{XRT_instrument}), \textit{Swift} Burst Alert Telescope ({\bat}; \citealt{BAT_instrument}) and \textit{Fermi} Large Area Telescope ({\lat}; \citealt{LAT_instrument}) observations overlap temporally. 

Several earlier works have already inferred spectral information on GRB~221009A in this energy window in temporal proximity. 
The {\gbm} data shows an afterglow component with an observed photon spectral index of $p_\gamma = -\dd \log N_\gamma / \dd \log E_\gamma \approx 2$ (40~keV–8~MeV), already shining through during low phases of the flaring prompt phase and turning more clearly into the afterglow after $\sim 700$~s \citep{FermiGBM_grb221009a, Zhang_GBMafterglow}. At 1.5~ks GRB~221009A gets occulted by the Earth and at 4~ks only an energy flux level of $EF_E \approx ~10^{-8} \,{\rm erg}/({\rm cm}^2~{\rm s})$ at 10~keV is estimated via the Earth occultation technique \citep{FermiGBM_grb221009a}.
\cite{swift_grb221009a} find preference for a break around 7~keV with photon spectral indices $p_\gamma$ of 1.7 (before the break) and 2.2 (after the break) from a combined fit of {\xrt}, {\bat} and the heavily absorbed {\uvot} data around 4.2~ks, when linking the spectral indices of a broken power law by 0.5. The residuals show a significant offset between {\xrt} and {\bat} fluxes, highlighting the challenges of the {\xrt} data in windowed timing (WT) mode (see Section~S1 of the supplementary materials for details). 
The photon spectral index $p_\gamma$ above the break seems to be in agreement with the rough value of $2$ determined from 3 energy bins (20-1220~keV, 5.1-25.7~ks) by \cite{konus}.
Earlier observations (1.35-1.86~ks) by GECAM-C (20~keV-6~MeV) show a photon spectral index of $p_\gamma \approx 2.1$, in agreement with their Earth occultation analysis (20-200~keV), although suggesting a harder value (1.6) above 200~keV \citep{AnEtAl}.
The combined analysis of {\xrt} and {\lat} in \cite{LaskarEtAl} suggests a break between the two instruments based on the time averaged photon spectral indices ($p_\gamma \approx 1.6-1.9$ for {\xrt}, $p_\gamma \approx 2.1$ for {\lat}), without specific information on the time interval or the data reduction in WT mode. 
Other works like \cite{LiuEtAl} and \cite{SternTkachev} also report photon spectral indices for {\lat} of approximately $p_\gamma \approx 2.1$. 

GRB~221009A gained particular attention due to its detection by {\lhaaso} at VHE energies \citep{lhaaso_grb221009a}. Significant detection by the Water Cherenkov Detector Array (WCDA) is however limited to less than 3~ks and upper limits are not significantly constraining for the time windows we consider here. A power-law extrapolation of the light curve to 4~ks indicates a flux at the level of a few $10^{-9} {\rm erg}/({\rm cm}^2~{\rm s})$ with big uncertainties of an order of magnitude arising from low statistics and EBL absorption. 
Due to poor observational conditions, in particular bright moonlight, no other VHE telescopes detected GRB~221009A.

Commonly, these observations are interpreted in a one-zone radiation model coupled to fireball blast wave dynamics \citep[for a review see e.g.][and references therein]{Zhang_GRBbook}.
Applying this to GRB~221009A required departures from this simple picture, in various directions \citep{RenEtAl,SatoEtAl,LaskarEtAl,GillGranot,OConnorEtAl,KannEtAl}. 

In order to clarify this emerging early afterglow spectrum --- namely, a hard spectrum (1.6) up to a break at a few keV followed by a consistent softer spectral index ($\gtrsim$2) up to 10~GeV --- and its compatibility with this one-zone standard model, we provide in this article a more comprehensive combined fit on the counts-level, using data from 0.6~keV up to 100~GeV.
In particular, we give a short summary of our reduced afterglow model, the data reduction and the joint-fit spectral analysis procedure in \autoref{sec:methods}, present and discuss our results in sections \ref{sec:results} and \ref{sec:discussion} and conclude in \autoref{sec:conclusion}.

\section{Methods} \label{sec:methods}

\subsection{Afterglow model}\label{sec:model}

The ultra-relativistic outflow of a GRB jet decelerates as it sweeps up material from the surrounding environment. A fraction of this swept up material is shock-accelerated to non-thermal energies and subsequently radiates away a part of its energy into the X-ray to gamma-ray energy bands. Our model is described in detail in \cite{Klinger_GRB190114C} and we give a short summary of the relevant parts here.

Following simple energy conservation arguments \citep{BlandfordMcKee1976}, we estimate the self-similarly decelerating bulk Lorentz factor $\Gamma$ of the outflow to be 34 (18) around 4~ks (22~ks), assuming an isotropic energy of $E_\mathrm{iso}=10^{55}$~erg \citep{FermiGBM_grb221009a} converted completely into radiation and a constant circum-burst density $n\st{up}=1/\mathrm{cm}^3$. We emphasise the weak dependence on these parameters $\Gamma\propto (E_\mathrm{iso}/n\st{up})^{1/8}$, which does not change dramatically for a wind-like density profile. We also assume that a fraction of the upstream ram pressure is converted to turbulent magnetic fields downstream (parameterised by $\varepsilon_B$, see \cite{Klinger_GRB190114C}).

During the particle acceleration process, the two competing energy loss processes for the electrons, namely adiabatic and radiative energy losses, collectively dictate the shape of the produced electron spectrum. We parameterise the latter as a smoothly broken power law with smoothness parameter $s_e$, and spectral indices based on the injected electron spectral index $p_e$ corresponding to the slow/fast cooling regime (see \cite{Klinger_GRB190114C} for the definitions of $\emin$, $\ecb$, $\emax$).
We note that we use the injection energy scale, $\emin$, as a free parameter, rather than the often adopted constant fraction parameters \citep[e.g.][]{SarietAl96}: the non-thermal particle number ($\zeta_e$) and energy injection ($\varepsilon_e$) rates \citep[Although see][]{VDHorstEtAl2014,MisraEtAl2021,WarrenEtAl2015,WarrenEtAl2018,ResslerLaskar2017,AsanoEtAl2020}. In the results section, however, the corresponding constraint on $\zeta_{e}$ and $\varepsilon_e$ are provided.
Furthermore, the parameter $\eta$ controls the electron acceleration time --- in units of the Bohm acceleration time --- which, in turn, dictates the maximum electron and thereby eventually also synchrotron and inverse Compton photon energy.

The quasi-steady-state energy distribution of these electrons subsequently dictates the spectral energy distribution (SED) of the non-thermal emission from the GRB. 
For the treatment of the radiative emission, we use the reduced synchrotron-self-Compton (SSC) model of \cite{Klinger_GRB190114C}, within which both a single-component synchrotron model (\textit{reduced syn}) as well as a two-component SSC model (\textit{reduced SSC}) can be treated as subclasses. It reduces all the degenerate complexity of the normalisation of the two components of the photon spectrum to two parameters: 1) the normalisation of the synchrotron component $F\st{syn}$ at an observed energy of 100~keV and 2) the relative factor of the inverse Compton component $N\st{IC}$, phenomenologically defined as the ratio of the maximum value of the energy flux spectra of both components. This source spectrum is then transformed to the observer's frame by assuming a point-like emission zone, an observation angle of $0^{\circ}$ and a redshift of $z=0.15$.

In summary, we explore with our \textit{reduced syn} model the parameters $p_e$, $s_e$, $\varepsilon_B$, $\emin$, $\eta$, $F\st{syn}$ and in case of the \textit{reduced SSC} model the additional parameter $N\st{IC}$ (for priors see Section~S4 of the supplementary materials). We fix  $E_\mathrm{iso}=10^{55}$~erg, $n\st{up}=1~\mathrm{cm}^{-3}$ and $t\st{obs}\approx 4.8 ~ (21.8)$~ks.

\subsection{Data reduction}
\subsubsection{{\xrt}}

The {\xrt} started observing GRB~221009A 3.3~ks after the \gbm{} trigger.  As reported
in \citet{swift_grb221009a}, the first 12 snapshots of XRT data, out
to 89~ks post-trigger, were obtained in Windowed Timing (WT) mode,
which is a fast (1.78~ms time resolution), 1D, readout mode, used to minimise the
effects of pile-up on the detected photons when a source is bright. 

1D spatial profiles
from the early WT observations indicated the presence of a strong dust
scattering component, which was later seen as a set of expanding
rings (i.e. dust scattering echos) in the 2D Photon Counting (PC) mode images.
Compared to a regular point source, these rings add a significant, but intrinsically inseparable component to the 1D region used to extract the GRB signal for the two time periods covered here ($3.9 - 4.5$~ks  and  $21.6 - 22.1$~ks; see Figure~S1).
By modelling the dust echo profile in WT mode, we have attempted to minimise its effect on the GRB spectral extractions and identified an appropriate background region for the later time interval (see Section~S1 of the supplementary materials for details).

In addition to the effect of the dust rings, we model the observed GRB flux suppression due to photoelectric absorption, whose exponential nature introduces particularly large uncertainties at lower energies ($\lesssim$ few keV). We use the conventionally adapted combination of an absorber close to the GRB and a small contribution of the Milky ways absorption ($N_\mathrm{H}=5.38\times 10^{21}\,\mathrm{cm}^{-2}$), as in \cite{swift_grb221009a}.

\subsubsection{{\bat}}
The {\bat} produces survey data when it is not triggering on a GRB.
As outlined in \cite{swift_grb221009a}, we used the BatAnalysis python package\footnote{This package is open source and is available on github at: https://github.com/parsotat/BatAnalysis } \citep{batanalysis} to download individual {\bat} survey data in the time windows 3907-4534s (4~ks) and 21.6-22.1~ks (22~ks) and process them to produce spectra in 8 energy bins from 14-195 keV (compare Section~S2 of the supplementary materials).

\subsubsection{{\lat}}
GRB~221009A became visible to {\lat} during orbital windows of around 2~ks, being occulted by the Earth for a bit more than 3~ks in-between.
The exact overlapping windows with {\xrt} and {\bat} result in very low statistic for {\lat}, such that we extend the temporal selection range for {\lat} to the maximal visible windows, 4-6~ks and 21.2-22.5~ks.

Due to the systematic effects from energy dispersion, we select only photons with energies above 200~MeV \citep{LAT_edisp}. We limit our energy range to 100~GeV, whereas the highest detected photon energies are 7.8~GeV (3~GeV) for 4~ks (22~ks).

Motivated by the large point spread function at low energies, we include photons within a cone of 10° around GRB~221009A. We highlight its distance of less than 5° from the Galactic plane means that the Galactic diffuse background emission is non-negligible on these timescales. This limits our ability to reliably infer the spectral index in the {\lat} energy range. We find that other background sources play no significant role (see Section~S3 of the supplementary materials for details).

\subsection{Joint-fit spectral analysis procedure}

The joint-fit spectral analysis is based on the method described in \cite{Klinger_GRB190114C}.
We define the likelihood $\mathcal{L}$ for each instrument on the counts level based on the corresponding suitable statistic: Poisson (4~ks) and {\tt C-stat}\footnote{Poisson data with Poisson background \citep{cash1979}} (22~ks) for {\xrt}, {\tt $\chi^2$} for {\bat}, and {\tt PG-stat}\footnote{Poisson data with Gaussian background} for {\lat}. The instrument response function is incorporated via forward folding. 
To account for inter-instrumental cross-calibration uncertainties we add a constant floating norm of up to $\pm15\%$ to each instrument \citep{Madsen2016,BAT_sys,LAT_sys}. To account for the expected large background contribution of the dust features of 10-20\% at 4~ks, we increase the range in this time window to a factor between 0.95 and 1.3 (see Section~S1 of the supplementary materials).

Following a Bayesian approach, we furthermore assume flat priors (see Section~S4 of the supplementary materials) to sample the parameters' posterior distributions and integrate the total evidence for model comparison via the Bayes factor.

\section{Results}  \label{sec:results}
We expect from single-instrument and combined-subset fits (see Section~S5 of the supplementary materials for details) that the SED should roughly resemble a broken power law.
This is based on a statistical preference for a break around 10~keV at 4~ks, while this is less clear for the 22~ks interval with lower statistics. At higher energies {\bat} and {\lat} can be fitted together by a power law without statistical preference for a break, with photon spectral indices of $\approx2.2$ (4~ks) and $\approx2.1$ (22~ks). We note that the large uncovered MeV gap from 0.2-200~MeV and the large uncertainties of {\lat}'s spectral index complicate the exclusion of additional features in-between.

We proceed with the results obtained from a combined fit including all three instruments, in order to extract the spectral properties of this rough broken-power-law behaviour in the framework of our \textit{reduced syn} model:
The break energy, the indices below and above the break and a potential additional spectral feature in between {\bat} and {\lat}.

We start by performing a combined fit using our \textit{reduced syn} model with a single component ($N_\mathrm{IC}=0$), our \textit{syn-only} case. 
Motivated by the non-standard {\xrt} background treatment, we perturb \textit{syn-only} case by releasing our assumption on the {\xrt} effective area correction (more than 30\%), the \textit{XRT-floating} case (\autoref{sec:XRTfloating}). 
Finally, we allow for an additional SSC component ($N_\mathrm{IC}>0$ in the \textit{reduced SSC}) in the \textit{SSC} case (\autoref{sec:SSC}).

\begin{figure*}
    \centering
    \includegraphics[width=0.95\linewidth]{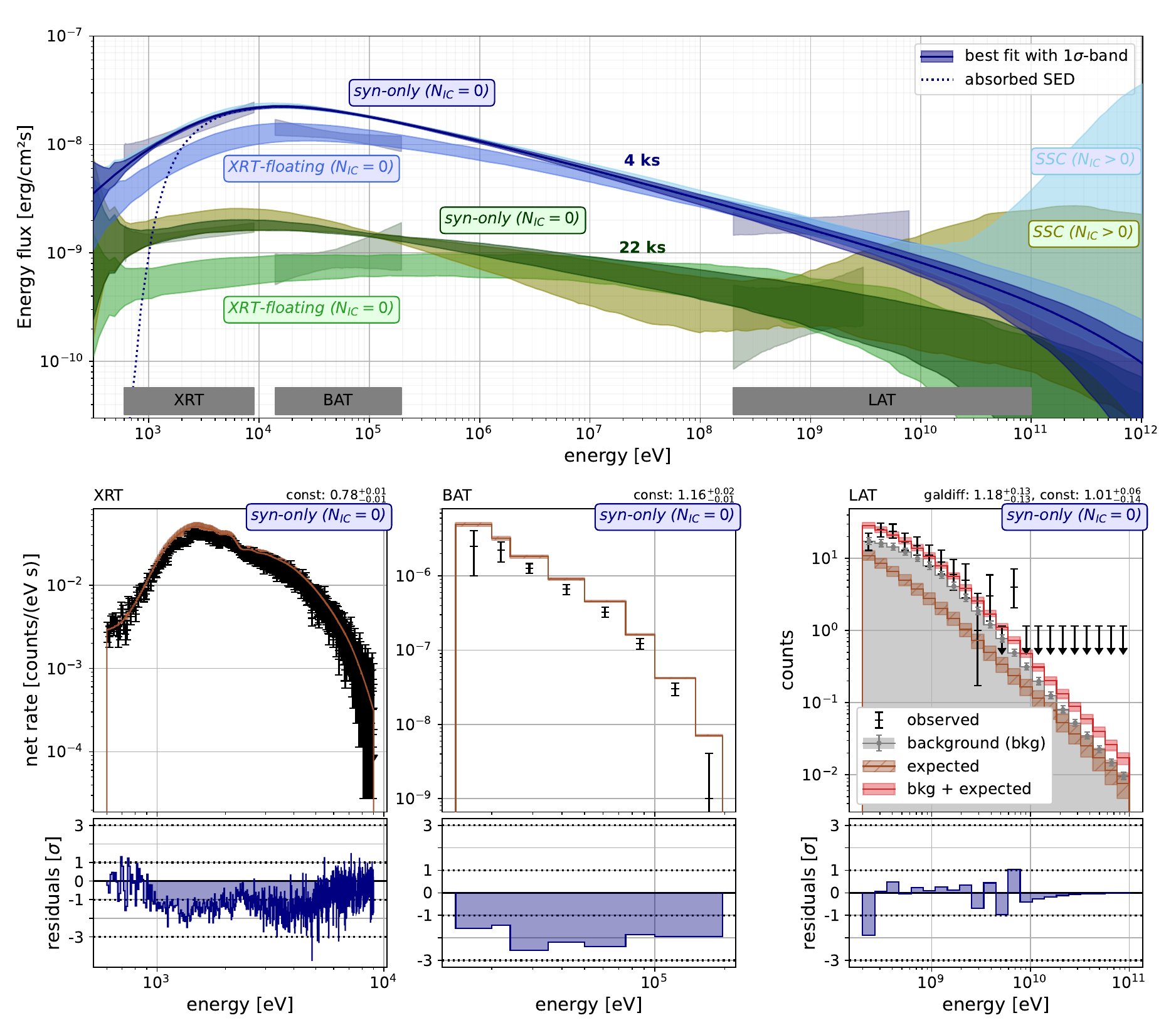}
    \caption{Top: $1\sigma$ envelope of intrinsic SED at 4~ks (blue) and 22~ks (green) for the cases: \textit{syn-only} ($N_{IC}=0$, darkest), \textit{XRT-floating} ($N_{IC}=0$ with extended floating norm factor for {\xrt}, lowest at 10~keV) and \textit{SSC} ($N_{IC}>0$, resulting in large uncertainties above 10~GeV). Grey butterflies correspond to single-instrument power law fits (compare Table~S5). The blue line corresponds to the best fit SED and the blue dashed one shows the observed, absorbed best fit spectrum (dotted line) for 4~ks interval. Bottom: Counts-level comparison for the \textit{syn-only}-case (blue dashed line from top panel) for each detector ({\xrt}, {\bat} and {\lat}) with corresponding residuals below. Floating norm factors for detectors ("const") and {\lat}'s galactic diffuse background template ("galdiff") correspond to default priors from \textit{syn-only}-case.
    \label{fig:result_t0}}
\end{figure*}

\autoref{fig:result_t0} shows for each time bin (4~ks in blue and 22~ks in green) the 3 best-fit models with a $1\sigma$-envelope and for the \textit{syn-only} case the corresponding residuals at the counts-level for all three detectors. The \textit{syn-only} case's posterior distributions are given in supplementary materials (Section~S6). The single instrument power law fits are added to the SED plot in grey to guide the eye.

The best fit shape of the \textit{syn-only} case resembles the broken power law expectation with a break energy of a few keV and a photon spectral index of $\approx2.2$.

\subsection{Perturbation 1: \textit{XRT-floating} case ($N_\mathrm{IC}=0$)}  \label{sec:XRTfloating}

Inspecting the residuals of {\xrt} and {\bat} (\autoref{fig:result_t0}) as well as the posterior distributions of their cross-correlation uncertainty factors (Figure~S2) indicates an increased effective area correction between them. 
The non-standard background treatment due to the dust ring pollution motivates us to explore in our \textit{XRT-floating} case the effect on our results when we extend the {\xrt} floating norm prior distribution to a shift of more than 30\% downwards. 
Such a preference for large effective area corrections of up to a factor of 2 is not uncommon for bright GRB afterglows (see e.g. \cite{Klinger_GRB190114C} for GRB~190114C). 

We observe for 4~ks that the {\xrt} data gets shifted down by a factor of $2\pm 0.3$, while the correction factors for {\bat} and {\lat} remain negligible ($\approx 1$). This results in a change of the spectral index above the break from $p_\gamma \approx 2.3 - 2.2$, corresponding to a change of the electron injection index from $p_e \approx 2.6 - 2.5$. We emphasise that the uncertainty to this values is likely underestimated due to the systematic effects in the {\xrt} analysis as well as the absorption model.
For 22~ks, {\xrt} requires a shift down by a factor of $2.2\pm 0.5$ and the evidence for a break is less clear than at 4~ks. Additionally, the posterior distributions become multimodal and very broad, such that the systematic uncertainties from the {\xrt} analysis are limiting the conclusions drawn from the \textit{syn-only} case at 22~ks. 

We conclude that an extended {\xrt} floating norm enlarges mainly the uncertainty on the inferred spectral index of the electrons.

\subsection{Perturbation 2: \textit{SSC} case ($N_\mathrm{IC}> 0$)}  \label{sec:SSC}
An additional spectral component is expected for the case $N_\mathrm{IC}> 0$, becoming relevant at the highest energies. The ad-hoc assumption of acceleration at the Bohm rate ($\eta=1$) results for both time intervals in a cut-off (the observed synchrotron burn-off) around a few GeV, where the {\lat} counts become sparse.
We investigate with this SSC perturbation the possibility that an IC component from an additional free parameter ($N_\mathrm{IC}>0$) might dominate the {\lat} observation by displacing the synchrotron cut-off to lower energies. 

We find no preference for this behaviour. Instead, the only difference to the \textit{syn-only} case's fit result is the additional uncertainty above a few GeV, see \autoref{fig:result_t0}. This visualises the limited power of the {\lat} data to constrain an IC component. 

We conclude that our combined fit is not sensitive to the onset of an additional IC component.

\section{Discussion}  \label{sec:discussion}

We discuss now the posterior distribution of the best fit scenarios, resulting at 4~ks in an apparent small uncertainty on the best fit values of $\varepsilon_B$, $p_e$ and $\emin$. We examine the validity of these uncertainties in the following.
In the \textit{reduced syn} model, both, the break energy and the spectral index below ($E_\gamma <10~$~keV) are driven by the combination of (i) the photoelectric absorption factor ($N_\mathrm{H}$ in proximity to the GRB) and (ii) the intrinsic break determined by the interplay between magnetic field ($\varepsilon_B \to$ cooling break energy) and minimum injected electron energy ($\emin$). Note that we directly treat $\emin$ as a fit parameter. 

Focusing on the intrinsic break energy within the \textit{reduced syn} model, the observed energy $\Egcbobs \approx$~\text{few}~keV can be matched by a break in the synchrotron component in three conceptually different solutions. They correspond to three different features in the electron spectrum: 1) \textit{slow cooling} ($\emin \ll \ecb $), i.e., a cooling break from the change from the dominance of adiabatic to synchrotron cooling; 2) \textit{fast cooling} ($\emin \gg \ecb $), i.e., a cooling tail in the electron spectrum or 3) their \textit{transition} ($\emin \approx \ecb $), i.e., a cut-off in the electron spectrum resulting in the low energy tail of the synchrotron spectrum of the lowest energy electrons transitioning to the synchrotron spectrum of the cooled electrons.

\begin{figure*}
    \centering
    \includegraphics[width=0.8\linewidth]{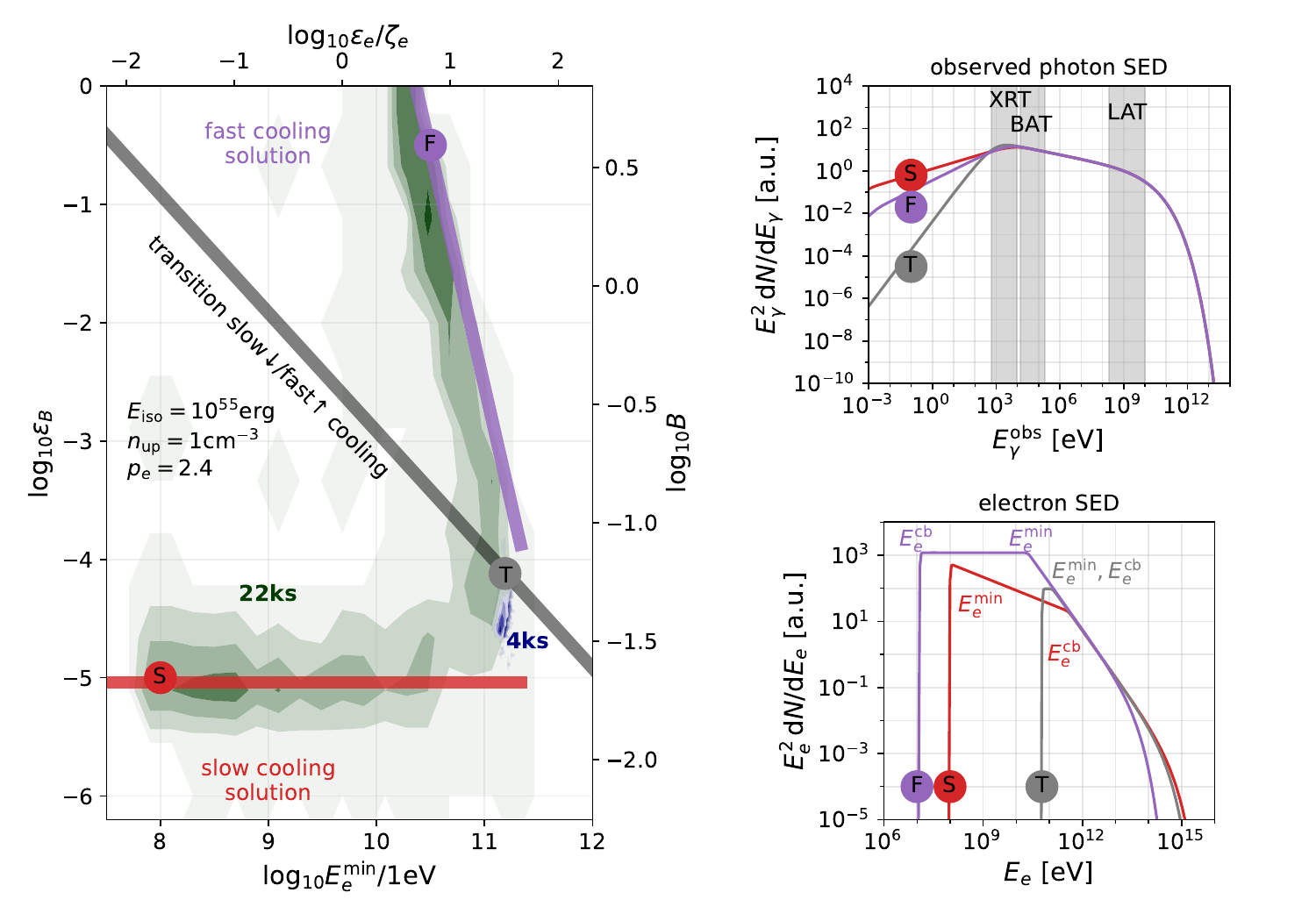}
    \caption{Left: posterior distributions in $\emin-\varepsilon_B$ plane for the 4~ks (small blue region) and 22~ks (large green boomerang) with the two arms fixing the observed break energy $\Egcbobs\sim 10$~keV, corresponding to the slow (red) and fast (purple) cooling solution, meeting in the transition region (grey). Note that we treat $\emin$ directly as a parameter instead of fixing it via $\varepsilon_e$ and our prior is $\emin > 10^8$~eV. For three representative solutions S (slow), F (fast) and T (transition) the differences in electron and observed photon spectral energy distributions are visualised.}
    \label{fig:fast_slow_cooling}
\end{figure*}

\autoref{fig:fast_slow_cooling} illustrates three representative cases for each regime, in terms of their electron (bottom) and photon (top) SEDs and shows the \textit{syn-only}-case's posterior distribution of the corresponding parameters ($\emin$ vs. $\varepsilon_B$, on the left). Placing $\Egcbobs\sim $~few~keV restricts the parameter space to the two arms (\textit{boomerang}), as is found for both time intervals (blue/green for 4~ks/22~ks). Note that we also provide a second x-axis and y-axis, to indicate the corresponding magnetic field strength  and ratio between number and energy of upstream electrons converted into non-thermal particles, respectively \footnote{$B=\Gamma\sqrt{32\pi \varepsilon_B n\st{up} m_p c^2 } $ and $\frac{\varepsilon_e}{\zeta_e} = \frac{(p_e-1)}{(p_e-2)}  \frac{\emin}{\Gamma m_p c^2} \approx \frac{\emin}{10\;\mathrm{GeV}}$}. 
We note that in particular the later mapping is based on multiple uncertain assumptions.

For all three solutions the photon spectrum above the break energy corresponds to the cooled electrons and is the same ($p_\gamma = -\dd \log N_\gamma / \dd \log E_\gamma = (p_e+1)/2$), whereas the spectral indices below differ: 4/3 for the transition solution, 1.5 for the fast cooling solution or $(p_e+1)/2 \approx 1.7$ (using an injected electron spectral index of $p_e=2.4$). Despite the power-law spectral index of $p_\gamma \approx1.6$ for {\xrt} alone, we find from our combined fits a strong preference in the data for a very hard spectrum below a few~keV ($p_\gamma \approx 0 $ from a broken-power law fit), which limits the best fit parameters on the \textit{boomerang} further into the lower-right corner, corresponding to the transition regime. 

It is important to note in this context the strong effect (one order of magnitude at 1~keV) of the exponential absorption factor in this spectral regime below the break energy, as visible in the dotted line in \autoref{fig:result_t0} showing the observed absorbed spectrum.

The interpretation of this hardening to discriminate between the three solutions is thus limited by the uncertainties of the absorption model, introduced by e.g. the chemical composition of the photoelectric model (see e.g. \citealt{DaltonMorris20_NHmetalicity,ValanetAl23_NHVariability}). For heavily absorbed sources with large statistics like GRB~221009A, it becomes relevant to propagate this uncertainty into the likelihood evaluation. A quantitative evaluation of these uncertainties is additionally complicated by the non-standard background estimation as well as uncertainties in the instrument response from e.g. instrumental edges and aging effects and goes beyond the scope of this paper. 

We thus conclude that an inferred limitation of the solution type to the transition regime in the 4~ks interval is not robust due to the underestimation of the above systematic uncertainties.

The inference of physical properties of the radiation zone from the observed break energy $\Egcbobs$, however, depends strongly on the solution type (qualitatively similar for wind-like density profiles):
\begin{equation}
    \Egcbobs \propto \begin{cases}
        \varepsilon_B^{-3/2} E\st{iso}^{-1/2} n\st{up}^{-1}  \hspace{1.5cm} & \text{slow cooling} \\
        \varepsilon_B^{1/2} E\st{iso}^{1/4} n\st{up}^{-1/4}  \left(\emin\right)^{2}, & \text{fast cooling} 
    \end{cases}
\end{equation}

Within the slow cooling solution, fixing the observed break position to the keV regime would effectively fix the comoving magnetic field to $B\approx 0.03\;\mathrm{G}$, $\varepsilon_{B}\to 10^{-5}$ (for $n\st{up}\approx 1$~cm$^{-3}$).
It is important to note that the steady state approximation is only accurate to a factor of order unity in this case.

On the other hand, the fast cooling solution is compatible with a large range of magnetic field values, up to a few Gauss. Here the dependence on the minimum injected electron energy $\emin$ becomes dominant, placing it from 10 to 100~GeV.
It is important to notice that in the fast cooling solution, in particular when approaching the transition solution, $\Egcbobs$ relates directly to the electron spectrum at the minimum injection scale, where the spectral shape is poorly understood. In particular the nature of the typically assumed step-like cut-on of the non-thermal electron spectrum is expected to be affected by the thermal population (see e.g. \citealt{Warren22_thermal}), but commonly neglected.

In terms of the micro-physical parameters, the left side of the slow cooling arm with $\varepsilon_e \ll \zeta_e$ requires the injection scale of the non-thermal electrons to be lower than the average proton energy (i.e. $\emin < (p-2)/(p-1) \Gamma m_p c^2 \approx 10$~GeV). This is in contrast to the typical values obtained from particle-in-cell simulations,  $\varepsilon_e \approx 3 \zeta_e$ \citep[ch.~4]{MarcowithEtAl2016}, a result which better aligns with the fast cooling solution we find ($\varepsilon_e/ \zeta_e \approx 3-10$). 
It is also worth mentioning that for the right side of the figure ($\emin \sim 100$~GeV), typical values of $\varepsilon_e \sim 0.1$ would require a value of $\zeta \sim 0.01$ (a percent of the electrons).
However, effects like pair-loading of the upstream medium can significantly influence this ratio, preventing us from drawing strong conclusions on the microphysics \citep{GroseljEtAl2022}.

Comparing our finding of $p_\gamma \approx 2.2-2.3$ to \cite{swift_grb221009a} and the observational picture sketched in \autoref{sec:intro}, we find consistency. Modeling results in other works find consistent values of the spectral index of the electrons, spanning $p_e \approx 2.2 - 2.8$, whereas the estimates for $\varepsilon_B \approx 10^{-1} - 10^{-7}$ vary greatly \citep{RenEtAl,SatoEtAl,LaskarEtAl,KannEtAl, GillGranot, lhaaso_grb221009a}. In particular our results for the electron spectral index $p_e$ help to constrain its value to 2.4--2.5.

Additionally, one could constrain a possible inverse Compton component using the {\lhaaso} non-observation at 4~ks to $N\st{IC} < 1$. 
We highlight, that our findings for GRB~221009A are thus also consistent with, but not more conclusive than the picture emerging from GRB~190114C \citep{Klinger_GRB190114C} and GRB~190829A \citep{HESS_190829A}, such that the physical origin of the VHE afterglow observations can not be determined more clearly from this brightest GRB detected so far.

\section{Conclusion} \label{sec:conclusion}

GRB~221009A was extraordinarily bright, which led to complications in the analysis of the data in several energy bands. Furthermore, its position close to the Galactic plane led to increased background levels which further broadened the systematic uncertainties.

Its SED from keV to GeV energies is in both time intervals (4~ks and 22~ks after $T_{0\mathrm{,GBM}}$) well described by a smoothly broken power-law, with break energy $\Egcbobs\approx$~few~keV. The photon spectral index above the break is ($p_\gamma = -\dd \log N_\gamma / \dd \log E_\gamma = 2.2-2.3$) with dominant systematic uncertainties driven by the {\xrt} background estimation. For the same reasons the photon spectral index below the break is less clear ($p_\gamma \approx 1.5-2$). 

Within our GRB spectral model, the inferred break energy $\Egcbobs\approx$~few~keV can have three possible interpretations. 
Either (1) as a slow cooling break, resulting in a weak magnetic field with only an upper limit on the minimum electron energy. Or (2) a fast cooling break, requiring a stronger magnetic field and a higher minimum electron energy. Or (3) the transition between (1) and (2), requiring a weak magnetic field and a large minimum electron energy. We summarise these results in Table~\ref{tab:results}. We note that the preference at 4~ks for the transition regime is not robust, and that both time intervals only limit $B$ and $\emin$ to the boomerang shape region shown in \autoref{fig:fast_slow_cooling}.

\begin{table}
\label{tab:results}
\begin{tabular}{l|l|l|l|l|l|l}
              & \vline & $B$ [G] & $\varepsilon_{B}$ & \vline & $\emin$ [eV] & $\varepsilon_{e}/\zeta_{e}$ \\ \hline
1) slow       & \vline & few  &  $10^{-5}$       & \vline &   $< 10^{11}$   & $<30$    \\
2) fast       & \vline & 0.1-few    &     $10^{-4}-1$              & \vline &   $10^{10}-10^{11}$      &   $3-30$    \\
3) transition & \vline &  0.1   &   $10^{-4}$                & \vline &   $10^{11}$      &   $30$                        
\end{tabular}
\caption{A summary of the constraints on key parameters for each of the three solutions of the energy break origin.}
\end{table}

\section*{Acknowledgements}
MK, AT and SZ acknowledge support from DESY (Zeuthen, Germany), a member of the Helmholtz Association HGF. The authors would also like to thank Michelle Tsirou for helpful input on the {\lat} analysis. This work was supported by the International Helmholtz-Weizmann Research School for Multimessenger Astronomy, largely funded through the Initiative and Networking Fund of the Helmholtz Association. SH acknowledges support from NSF grant AST2205917. All figures by the authors under a \href{https://creativecommons.org/licenses/by/4.0/}{CC BY 4.0 license}.

\section*{Data Availability}
The X-ray data underlying this paper are publicly available from the UK {\em Swift}
archive at \url{https://www.swift.ac.uk/swift_live/} or the HEASARC
archive at \url{https://heasarc.gsfc.nasa.gov/cgi-bin/W3Browse/swift.pl}. {\lat} data is available from \url{https://fermi.gsfc.nasa.gov/ssc/data/access/}.
The fits have been performed with the publicly available Multi-Mission Maximum Likelihood framework \citep[3ML;][]{3ML_ref} and the UltraNest package \citep[][]{Ultranest_ref}. The code of the authors will be made available upon request.



\bibliographystyle{mnras}
\bibliography{references} 








\bsp	
\label{lastpage}
\end{document}


\label{firstpage}
\pagerange{\pageref{firstpage}--\pageref{lastpage}}
\maketitle



\section{Details on {\xrt} data reduction} \label{ap:xrt}

The XRT analysis is complicated by the presence of prominent dust echos. These are caused by the scattering of X-rays from the bright,
GRB prompt emission back into the XRT field-of-view by Galactic dust density enhancements along the line-of-sight. 

To mitigate the effect of dust scattering on the data, we modeled the 1D spatial dust echo profiles of the two WT observations. To this end, we adopted the dust column density distribution along the line of sight from Fig.~4 of \citet{swift_grb221009a}. We then reconstructed the 2-dimensional  dust echo at the mid-times of the two WT observations of the afterglow, using the Mathis, Rumpl and Norsieck dust model {\citep{mathis:77}} adopted in \citet{swift_grb221009a}, as implemented in our dust scattering code {\tt dscat} \citep{heinz:16}. 

The radial intensity profiles derived from the dust distribution intensities were then projected onto the XRT focal plane, multiplied by the XRT effective area, and vignetted. This procedure generated what would have been observed as XRT images of dust rings in PC mode at the time of the two WT observations. These images were masked over a $600\times600$ pixel grid at the 2.357~arcsec spatial scale of the XRT CCD detector, and then summed in the vertical direction to mimic the WT readout process of the 1D WT dust scattering contribution.

\begin{figure*}
    \centering
    \includegraphics[width=0.95\linewidth]{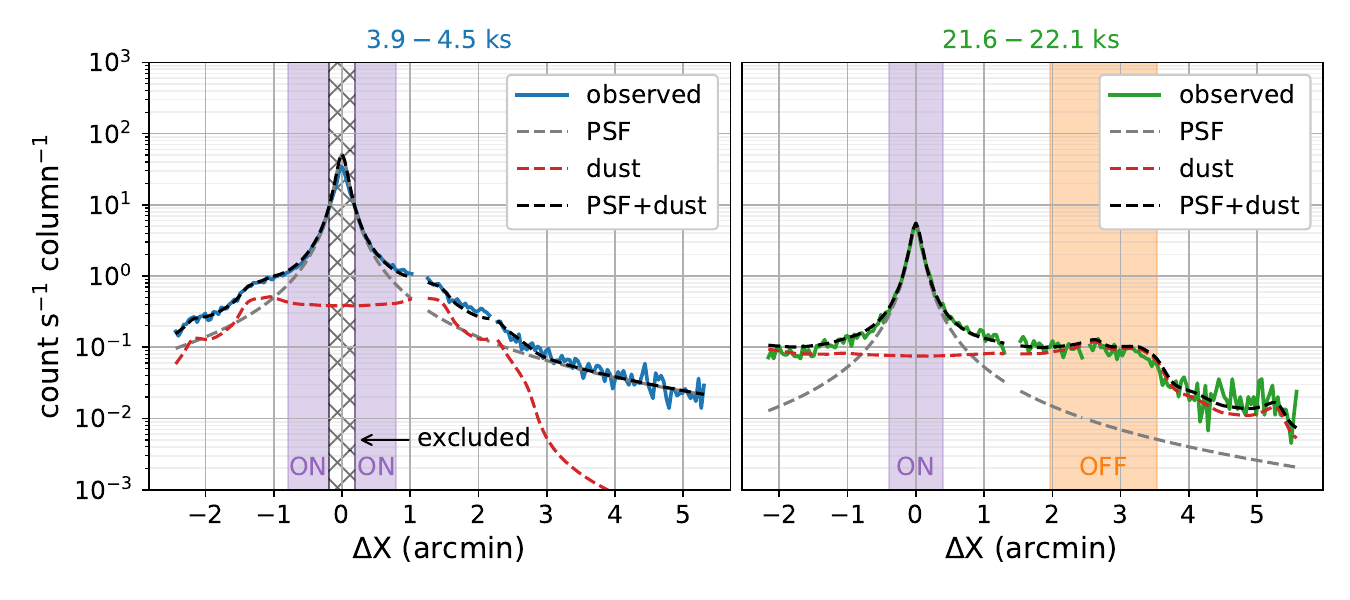}
    
    \caption{Observed {\xrt} 1D WT profiles created over the $0.8-5.0$~keV band (blue: 4~ks; green: 22~ks), along with predicted profiles from the central point source (grey), the dust scattered rings \citep[red; derived from the dust distribution modelling presented in][]{swift_grb221009a} and the total model (black). The left panel shows the data and model
    at 4~ks, with inner and outer source extraction region
    radii delineated by the shaded region labelled `ON'. The right panel shows the equivalent results at 22~ks,
    with an additional background region identified by the shaded region labelled `OFF'.}
    \label{fig:wt_profiles}
\end{figure*}

To demonstrate the fidelity of this approach, we plotted these dust profiles and the expected central point source profile in \autoref{fig:wt_profiles}). 

Given the initial, intense X-ray emission from the GRB, the WT data at 4~ks are piled-up, the effect of which is to suppress the observed counts in the core of the point spread function and harden the spectrum. To accounted for this, an inner extraction region radius of 11.8~arcsec was used to exclude counts from the piled-up core (cf. grey-crossed region in \autoref{fig:wt_profiles}).

The outer extraction radius was limited to 47.1~arcsec in order to minimise the contribution of dust scattered emission in the extraction region. The XRTDAS software task {\sc xrtmkarf} was used to adjust the on-axis XRT effective area for the expected fractional PSF losses encountered with this extraction region.
The profile modelling shown in \autoref{fig:wt_profiles} indicates the dust contributes 15\% of the counts in the extraction region, which needs to be accounted for in the cross-normalisation constant when spectral fitting with the higher energy instruments not affected by dust.

While the GRB emission is no longer bright enough to cause pile-up in
the WT data obtained at 22~ks, the relative contribution from the dust
scattered emission is now higher. The profile modelling in
\autoref{fig:wt_profiles} shows the dust echo provides an
approximately flat background level out to 3.54~arcmin radius, which
then drops by a factor of six. This radius corresponds to the dense
dust cloud at a distance of $400-600$~pc, identified by \citet{swift_grb221009a}.
Given the nature of the profile at this time, we limited the source extraction region to a radius of 23.6~arcsec and used a background region from $1.96-3.54$~arcmin, as shown
in \autoref{fig:wt_profiles}, then updated the source and background spectral file {\sc backscal}
keywords appropriately.\footnote{See \url{https://www.swift.ac.uk/analysis/xrt/backscal.php}}

The xrtmkarf correction is accurate to $\sim 10$\% nominally (when no pile-up corrections are involved), but could increase to $\sim 20-25$\% when extreme pile-up corrections are required. The BAT-XRT cross-calibration normalisation is typically
good to $\sim 10-20$\%, but extreme outliers have been seen ($\sim 30-40\%$).

In addition to the dust ring contribution of the emission scattered on the dust clouds, we considered along the line-of-sight the commonly used galactic (``TBabs'') and extragalactic (``zTBabs'') photoelectric absorption models, where the abundance and cross-section are set to ``wilm'' \citep{Wilms2000} and ``vern'' \citep{Verner1996}, respectively. However, we emphasise the unparameterised uncertainty of this model introduced by the assumptions on the abundances.

\section{Details on {\bat} data reduction} \label{ap:bat}
The survey data that was analyzed here correspond to observation IDs 01126854000 
 (4~ks) and 01126853001 (22~ks). The energy bins are 14-20, 20-24, 24-35, 35-50, 50-75, 75-100, 100-150, and 150-195 keV.

\section{Details on {\lat} data reduction} \label{ap:lat}
We use 3ML's \texttt{FermiPyLike} plugin for our fits.
We model the galactic and isotropic diffuse background emission with the latest templates provided by the {\lat} collaboration, {\tt gll\_iem\_v07} and {\tt iso\_P8R3\_TRANSIENT020E\_V3\_v1}, respectively. We allow for a systematic uncertainty on the galactic diffuse background template, which varies around 10-20\% in our fits. Despite its proximity to the galactic plane, we find neighbouring sources to have a count contribution on the percent-level compared to the galactic diffuse background and include these effects into the floating norm of the galactic template.

\section{Details on Joint-fit spectral analysis} \label{ap:jointfit_analysis}
We use a {\tt python} package, the Multi-Mission Maximum Likelihood framework \citep[3ML;][]{3ML_ref}, for our analysis.
We assume uniform ($p_e \in [1.5, 3]$) and log-uniform ($\varepsilon_B\in[10^{-9}, 1]$, $\emin\in[10^{8}, 10^{15}] \; \mathrm{eV}$, $\eta\in[10^{-3}, 10^{3}]$, $F\st{syn}\in[10^{-10}, 10^{-7}] \; \mathrm{erg}\mathrm{cm}^{-2}\mathrm{s}^{-1}$, electron break smoothness $s\in[10^{-0.5}, 10]$,  and $N\st{IC}\in[10^{-4}, 10]$) 
priors $\pi$. Calling data $\mathcal{D}$, model $\mathcal{M}$, parameter vector $\vec{\theta}$ and likelihood $\mathcal{L}$, we derive the posterior probability distribution and the evidence $Z$ \citep[e.g.][]{KassRaftery95,Trotta2008}
\begin{eqnarray}
    Z = \int \dd \vec{\theta}  \; \mathcal{L}(\mathcal{D}|\vec{\theta},\mathcal{M}) \: \pi(\vec{\theta},\mathcal{M}) \:,
\end{eqnarray}
with the nested sampling Monte Carlo algorithm MLFriends, implemented in the {\tt python} package UltraNest \citep{Ultranest_ref}.

\section{Phenomenological picture from subsets of instruments} \label{ap:PL_fits}

For both time intervals, 4~ks, and 22~ks, we performed fits to each instrument alone using a power law to get an intuition for the spectral index in this energy band. The results are summarised in \autoref{tab:fitParams} and plotted as grey envelopes in the SEDs of Figure~1. 

No instrument shows preference for a spectral break, with limited robustness for different reasons for each. 
For {\xrt} the fit includes the exponential absorption term with step features from the photoelectric cross section, which hinders clear identification of spectral features. 
Since {\bat} has only 8 energy bins, lower count rates in the lowest/highest bin suggest curvature, however these bins also suffer from highest uncertainties. 
The {\lat} counts are strongly affected by the galactic background emission and in the highest energy bins the low counts ($\leq 3$) introduce large uncertainties. This prevents robust inference of the spectral index and limits the {\lat} data alone to information on the energy flux level.

Combined power-law vs. smoothly-broken-power-law fits of {\xrt} and {\bat} require at 4~ks a break at an energy between 5 and 10 keV, depending on the prior width of the floating norms ($\Delta \log_{10}Z \approx 50$). At 22~ks, either a break or an extended floating norm of around a factor $2.4 \pm 0.5$ statistically only indicate a minor improvement at a similar level ($\Delta \log_{10}Z \approx 3$).

{\bat} and {\lat} data combined can be fit well by a single power law for both time intervals, see also \autoref{tab:fitParams}.

A combined fit to all three data sets with a smoothly broken power law confirms this picture of a break in the keV regime and a single power law component extending up to the GeV energies.

\begin{table} 
\begin{tabular}{|l|l|l|l|l|l|}
\hline
dataset                  & time & spectral index $p_\gamma$ & energy flux $E_FE|_{E_0}$ at $E_0$ {[}erg/(cm²s){]} & $E_0$ {[}eV{]}        & $N_H [10^{22}/\mathrm{cm}^2]$ \\ \hline
\multirow{2}{*}{\xrt}     & 4ks           & $1.64 \pm 0.01$   & $(1.89 \pm 0.01)\times 10^{-8}$    & $4\times 10^{3}$   & $1.28 \pm 0.03$   \\ \cline{2-6} 
                         & 22ks          & $1.88 \pm 0.04$   & $(1.56 \pm 0.02)\times 10^{-9}$    & $4\times 10^{3}$   & $1.51 \pm 0.06$   \\ \hline
\multirow{2}{*}{\bat}     & 4ks           & $2.1 \pm 0.1$     & $(1.3 \pm 0.1)\times 10^{-8}$      & $3.1\times 10^{4}$ &                   \\ \cline{2-6} 
                         & 22ks          & $1.8 \pm 0.3$     & $(8 \pm 2)\times 10^{-10}$         & $3.1\times 10^{4}$ &                   \\ \hline
\multirow{2}{*}{\lat}     & 4ks           & $2 \pm 0.2$       & $(1.8 \pm 0.3)\times 10^{-9}$      & $1.6\times 10^{9}$ &                   \\ \cline{2-6} 
                         & 22ks          & $1.8 \pm 0.4$     & $(3.6 \pm 1.6)\times 10^{-10}$     & $1.6\times 10^{9}$ &                   \\ \hline
\multirow{2}{*}{\bat+\lat} & 4ks           & $2.2 \pm 0.02$    & $(1.29 \pm 0.04)\times 10^{-8}$    & $3.1\times 10^{4}$ &                   \\ \cline{2-6} 
                         & 22ks          & $2.07 \pm 0.05$   & $(8 \pm 2)\times 10^{-10}$         & $3.1\times 10^{4}$ &                   \\ \hline
\end{tabular}
    \caption{Individual instrument fit parameters to a power law model $EF_E = E_FE|_{E_0}  (E_\gamma/E_0)^{p_\gamma} $ with two additional photoelectric absorption terms \texttt{TbAbs} and \texttt{zTbAbs} for {\xrt}. Uncertainties correspond to the 1$\sigma$ interval. }\label{tab:fitParams}
\end{table}

\section{Posterior distributions } \label{ap:posteriors}

\begin{figure}
    \centering
    \includegraphics[width=\linewidth]{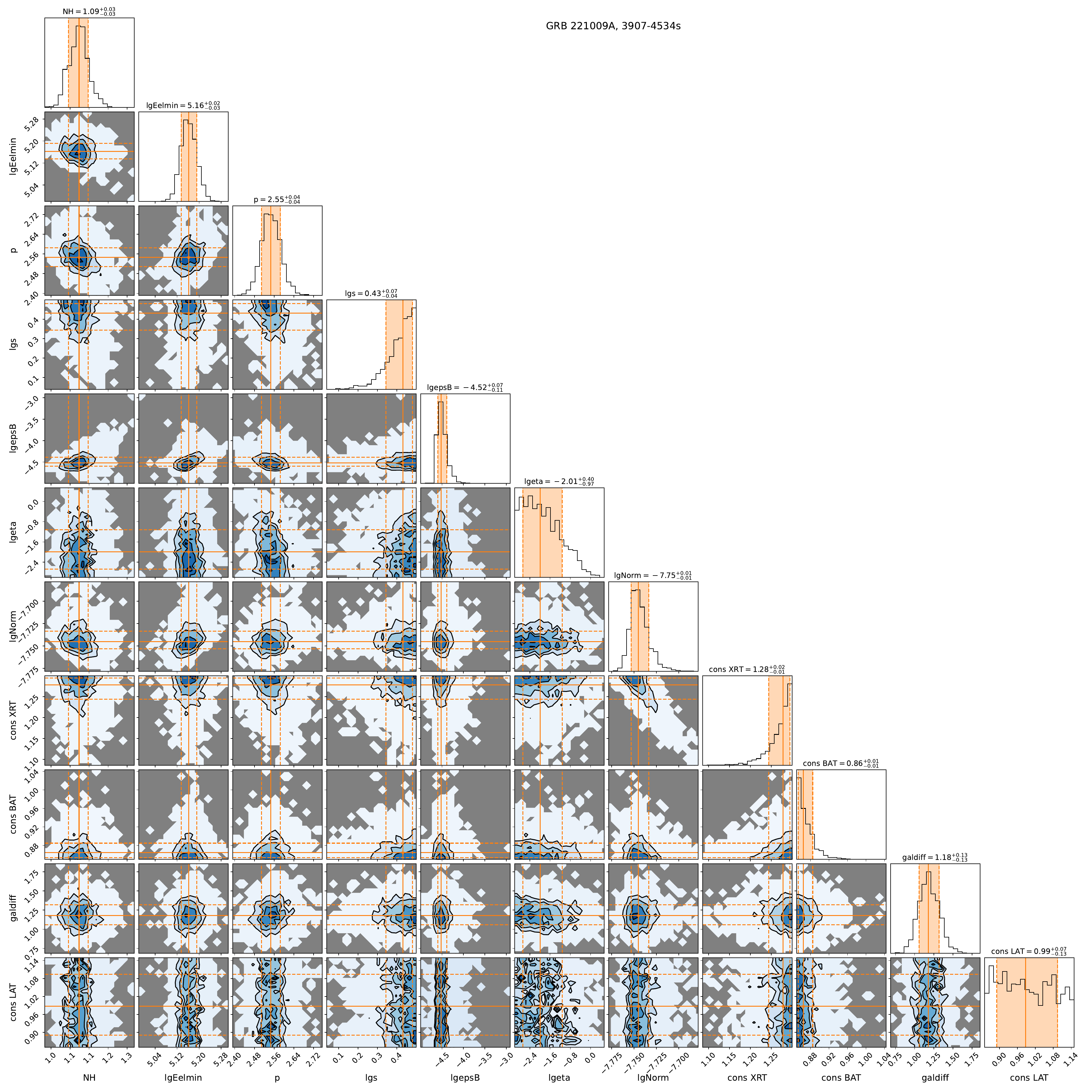}
    \caption{Posterior distributions for the \textit{syn-only}-case for the 4~ks time interval. NH is the column density of the photoelectric absorption in $10^{22}$~atoms~per~cm$^2$, $p=p_e$ is the spectral index of the injected electrons that cool to a break of smoothness $s_e=10^\mathrm{lgs}$,$\emin = 10^\mathrm{lgEelmin}$~MeV is the minimum injected electron energy, $\varepsilon_B = 10^\mathrm{lgepsB}$ defines the magnetic field strength, $\eta= 10^\mathrm{lgeta}$ is inversely proportional to the maximum energy of the electrons and thus also photons and $ F\st{syn}= 10^\mathrm{lgNorm}$ is the reference value for the synchrotron photon flux in $erg/(cm^2s)$ at $\Eg\ut{obs} = 10$~keV. cons\_XRT, cons\_BAT, cons\_LAT are the floating norm factors and galdiff is the floating norm of the galactic diffuse background template.}
    \label{fig:corner_syn_default_t0}
\end{figure}



\bibliographystyle{mnras}
\bibliography{references} 





\bsp	
\label{lastpage}